\documentclass[aps,10pt,prb, twocolumn, superscriptaddress]{revtex4-2}
\usepackage{bm, amsmath, amsfonts, amssymb, mathtools, braket}
\usepackage{multirow}
\usepackage{graphicx}
\usepackage{float, xcolor}
\usepackage{physics}
\usepackage{comment}

\usepackage[
pagebackref=false,
colorlinks=true,
linkcolor=blue,
urlcolor=blue,
filecolor=black,
citecolor=blue,
pdfstartview=FitV,
pdftitle={},
pdfauthor={},
pdfsubject={},
pdfkeywords={},
bookmarksopen=true
]{hyperref}

\newcommand{\1}{\mbox{1}\hspace{-0.25em}\mbox{l}}

\def\re{\text{Re}}
\def\im{\text{Im}}

\def\sgn{\text{sgn}}

\begin{document}
\title{
Winding Topology of Multifold Exceptional Points
}

\author{Tsuneya Yoshida}\email{yoshida.tsuneya.2z@kyoto-u.ac.jp}
\affiliation{Department of Physics, Kyoto University, Kyoto 606-8502, Japan Kyoto, Japan}
\affiliation{Institute for Theoretical Physics, ETH Zurich, 8093 Zurich, Switzerland}
\author{J. Lukas K. K\"onig}\email{lukas.konig@fysik.su.se}
\affiliation{Department of Physics, Stockholm University, AlbaNova University Center, 10691 Stockholm, Sweden}
\author{Lukas R\o dland}\email{lukas.rodland@fysik.su.se}
\affiliation{Department of Physics, Stockholm University, AlbaNova University Center, 10691 Stockholm, Sweden}
\author{Emil J. Bergholtz}\email{emil.bergholtz@fysik.su.se}
\affiliation{Department of Physics, Stockholm University, AlbaNova University Center, 10691 Stockholm, Sweden}
\author{Marcus St\aa lhammar}\email{marcus.backlund@physics.uu.se}
\affiliation{Nordita, KTH Royal Institute of Technology and Stockholm University, Hannes Alfv\'ens v\"ag 12, SE-106 91 Stockholm, Sweden}
\affiliation{Institute for Theoretical Physics, Utrecht University, Princetonplein 5, 3584CC Utrecht, The Netherlands} 
\affiliation{Department of Physics and Astronomy, Uppsala University, Uppsala, Sweden}

\date{\today}

\begin{abstract} 
Despite their ubiquity, a systematic classification of multifold exceptional points, $n$-fold spectral degeneracies (EP$n$s), remains a significant unsolved problem.
In this article, we characterize
the Abelian eigenvalue topology of generic EP$n$s and symmetry-protected EP$n$s for arbitrary $n$. The former and the latter emerge in a $(2n-2)$- and $(n-1)$-dimensional parameter space, respectively.
By introducing topological invariants called resultant winding numbers, we elucidate that these EP$n$s are stable due to topology of a map from a base space (momentum or parameter space) to a sphere defined by resultants. 
In a $D$-dimensional parameter space ($D\geq c$),
the resultant winding number topologically characterize
a $(D-c)$-dimensional manifold of generic [symmetry-protected] EP$n$s whose codimension is $c=2n-2$ [$c=n-1$].
Our framework implies fundamental doubling theorems for both generic EP$n$s and symmetry-protected EP$n$s in $n$-band models.
\end{abstract}

\maketitle

\section{Introduction}
\label{sec: intro}
The discovery of topological semimetals is one of the significant successes in modern condensed matter physics. A prime example is a Weyl semimetal~\cite{Wan_WSM_RB2011,Burkov_WSM_PRL2011,Silaev_WSMSuper_PRB2012,Turner_WSM_Textbook2013,Yan_WSMReview_AnnRevCondMattPhys2017} 
realized in $\mathrm{TaAs}$~\cite{Xu_WSMTaAsExp_Science2015,Lv_WSMTaAsExp_PRX2015,Huang_TaAsNegMagReg_PRX2015} which hosts elementary excitations with linear dispersion analogous to Weyl fermions~\cite{Weyl_WeylFermi_ZPhys1929}.  
For the emergence of Weyl fermions in solids~\cite{Xu_WSMTaAsExp_Science2015,Lv_WSMTaAsExp_PRX2015,Huang_TaAsNegMagReg_PRX2015,Yang_WSMMn3SnLDA_NJP2017,Higo_WSMMn3SnExp_APL2018,Matsuda_WSMMn3SnQAH_NatComm2020}, topologically protected two-band touching in three dimensions plays an essential role which induces negative magnetoresistance. Remarkably, considering more than two bands further enriches the topological semimetals and allows for exotic particles that have no high-energy analogue~\cite{Manes_MFermi_PRB2012,Bradlyn_MFermi_Science2016,Weng_MFermi_PRB2016,Chang_MFermiRhSi_PRL2017,Tang_MFermi_PRL2017}.
Specifically, multifold fermions emerge which are described by topologically protected multiband touching points.
These particles are extensively studied~\cite{Flicker_MFermiOptRes_PRB2018,Martinez_MFermiLinOpt_PRB2019,Maulana_MFermiRhSi_PRR2020,Gao_MFermiSuper_QFront2022,Hsu_MFermiQGeo_PRB2023,Balduini_MFermiExp_Nature2024} and three-, four-, and six-fold fermions are realized in materials so far~\cite{Rao_3fldFermiCoSi_Nature2019,Takane_3and4fldFermiCoSi_PRL2019,Schroter_6fldFermiAlPt_NatPhys2019,Sanchez_4fldFermiRhSi_Nature2019}.

Recently, studies on non-Hermitian systems have opened up a new research direction of topological physics~\cite{Hu_nH_PRB11,Esaki_nH_PRB11,Hatano_PRL96,Hatano_nHSkin_PRB1997,TELeePRL16_Half_quantized,ZPGong_PRL17,Alvarez_nHSkin_PRB18,SYao_nHSkin-1D_PRL18,SYao_nHSkin-2D_PRL18,KFlore_nHSkin_PRL18,Yokomizo_BBC_PRL19,Lee_SkinPRB19,Borgnia_ptGapPRL2020,Zirnstein_ptGapPRL2021,Zhang_BECskin19,Okuma_BECskinPRL19,Song_LSkin_PRL2019,Xiao_nHSkin_Exp_NatPhys20,Gong_class_PRX18,KKawabata_TopoUni_NatComm19,Kawabata_gapped_PRX19,Zhou_gapped_class_PRB19,Bergholtz_Review19,Ashida_nHReview_AdvPhys,Okuma_NHReview_AnnRevCondMatt2023,Lin_NHSEReview2023}.
One of the unique phenomena is the emergence of exceptional points (EPs) which are band-touching points of both real and imaginary parts accompanied by the coalescence of eigenvectors~\cite{TKato_EP_book1966}. Such non-Hermitian band touching is stable in two dimensions and is protected by non-Hermitian topology which is characterized by winding of complex energy eigenvalues~\cite{HShen2017_non-Hermi,Kawabata_gapless_PRL19,Yang_EPdoubling_PRL2021}, i.e., Abelian topology.
Considering multiband also enriches EPs~\cite{Demange_EP3MathPhoto_JPhysA2012,Delplace_Resul_PRL21,Mandal_EP3_PRL21,Sayyad_EPn_PRR2022,Konig_BraidEP3PRR2023,Yang_EPhomoto_arXiv2023,Yang_EPn_PRB2023,Montag_EPn_PRR2024}. Generic $n$-fold EPs (EP$n$s) in the absence of symmetry emerge in $2n-2$ dimensions; they require tuning $2n-2$ parameters~\cite{Rotter_EP_JPA09,Berry_EP_CzeJPhys2004,Heiss_EP_JPA12,HShen2017_non-Hermi,YXuPRL17_exceptional_ring,VKozii_nH_arXiv17,KTakata_EP_PRL2018,Zyuzin_nHEP_PRB18,Yoshida_EP_DMFT_PRB18,Yang_EPdoubling_PRL2021,Yoshida_EPDiscInd_PRB2022}, which is a strikingly small number given that generic $n$-fold Hermitian band touching requires $n^2-1$ tuning parameters~\footnote{
At $n$-fold band-touching point of Hermitian cases, the $n\times n$ Hamiltonian should be proportional to the identity matrix, which leads to $n^2-1$ constraints.
We note that $n\times n$ Hamiltonians are described by $n^2$ real numbers. The diagonal elements are described by $n$ real numbers. The off-diagonal elements are described by $n^2-n$ real numbers because they are complex but related to each other because of Hermiticity. Therefore, $n^2-1$ parameters need to be tuned for $n$-fold band touching\!
}. 
In the presence of parity-time ($PT$) symmetry, this number is further reduced to $n-1$~\cite{Budich_SPERs_PRB19,Okugawa_SPERs_PRB19,Yoshida_SPERs_PRB19,Zhou_SPERs_Optica19,Kawabata_gapless_PRL19,Yoshida_SPERs_mech19,Kimura_SPES_PRB2019,Yoshida_nHReview_PTEP20,Stalhammar2021,Sayyad2023}.
In addition, EP$n$s in multiband systems may exhibit non-Abelian topology~\cite{Konig_BraidEP3PRR2023,Yang_EPhomoto_arXiv2023} [e.g., non-Abelian charge (braiding) of EP2s in systems with three or more bands]
The emergence of EP$n$s is ubiquitous and has been reported for a variety of systems~\cite{Lin_EP3phtonic_RPL2016,Schnabel_EP3PTPhotonic_PRA2017,Wiersig_EP5photo_PRA2022,EP3_LiebLatt_PRB2020,Delplace_Resul_PRL21,Hatano_EP3Lindblad_MolPhys2019,Khandelwal_EP3Lindblad_PRXQ2021,Crippa_EP4Corr_PRB2021,Gohsrich_EPnHN_arXiv2024,Wu_EP3openQ_NatNano2024,Bai_NLEP3_PRL2023,Liu_EPn_PRR2023,Tang_EPn_Science2020,Tang_EP3Mech_NatComm2023}, such as photonic systems~
\cite{Lin_EP3phtonic_RPL2016,Schnabel_EP3PTPhotonic_PRA2017,Wiersig_EP5photo_PRA2022,EP3_LiebLatt_PRB2020,Wang_EPnPTPhotoSciAdv2023,Hodaei_EPnPT_Nature2017} and open quantum systems~\cite{Hatano_EP3Lindblad_MolPhys2019,Khandelwal_EP3Lindblad_PRXQ2021} and so on.

Despite their ubiquity, systematic topological characterization of EP$n$s remains a crucial unsolved issue; only symmetry-protected EP3s are characterized so far~\cite{Delplace_Resul_PRL21,Tang_EP3Mech_NatComm2023}. One theoretical challenge lies in missing topological invariants characterizing EP$n$s for arbitrary $n$ which quantify their robustness.

In this paper, we systematically characterize the Abelian topology of both generic and symmetry-protected EP$n$s [see Table~\ref{tab: W}]. 
Through a proper choice of resultants, we 
generalize resultant winding numbers~\cite{Delplace_Resul_PRL21}. 
In a $D$-dimensional parameter space with $D\geq c$,
our resultant winding number characterize a $(D-c)$-dimensional manifold of generic [symmetry-protected] EP$n$s, whose codimension is $c=2n-2$ [$c=n-1$].
The characterization of the symmetry-protected EP$n$ is straightforwardly extended to systems with pseudo-Hermiticity, charge-conjugation-parity ($CP$), or chiral symmetry. Furthermore, the introduced resultant winding numbers lead to doubling theorems of EP$n$s in $n$-band systems.

The rest of this paper is organized as follows.
In Sec.~\ref{sec: EP3 4D}, we illustrate our argument of the topological characterization and the doubling theorem for generic EP3s in four dimensions.
Generalizing this argument, we 
address the topological characterization of EP$n$ for arbitrary $n$; Secs.~\ref{sec: EPn nosym W} and~\ref{sec: EPn PTsymm} address the case of generic EP$n$ and symmetry-protected EP$n$, respectively.
Sec.~\ref{sec: summary} provides a short summary and outlook. 
The Supplemental Material (SM) is devoted to technical details of resultants; we show equivalence between zero resultant vectors and EP$n$s in SM Sec.~\ref{sec: R=0 vs EPn}, 
compute the resultant winding number for an acoustic system in SM Sec.~\ref{sec: EP3 phonon},
provide a generic toy model for EP$n$s in SM Sec.~\ref{sec: nxn toy model}, 
and explicitly calculate the resultant winding of models with an arbitrary winding number in SM Sec.~\ref{sec: EPn nosymm W=m}.

\begin{table}[b]
    \centering
    \begin{tabular}{cccc}
          \hline\hline
       ~Symmetry~   & ~$c$~               & ~Winding Number~ & ~Resultants $\bm{R}$~                                                                                \\      \hline 
	      none          &	$2n-2$	   & Eq.~\eqref{eq: W(2n-3) nosymm}         &  Eq.~\eqref{eq: Rvec NoSym 2n-2}                                              	       \\
	$PT$ psH        &	\multirow{2}{*}{$n-1$}	&    \multirow{2}{*}{Eq.~\eqref{eq: W PT}}  	& \multirow{2}{*}{ Eq.~\eqref{eq: Rvec PT} with $H'$ }      	       \\ 
	$CP$ CS  	       \\ \hline\hline
    \end{tabular}
     \caption{ 
     EP$n$s in $c$ dimensions and the winding numbers characterizing their Abelian topology.
     ``$PT$" and ``$CP$" denote parity-time symmetry and charge-conjugation-parity symmetry, respectively. ``psH" and ``CS" denote pseudo-Hermiticity and chiral symmetry, respectively. 
     A generic (symmetry-protected) EP$n$ emerges in $c=2n-2$ ($c=n-1$) dimensions.       
     These EP$n$s are characterized by winding numbers of resultants which are specified by the third and fourth columns. In the presence of symmetry, resultants are computed from the characteristic polynomial $P(\lambda)=\mathrm{det}[H'-\lambda \1]$ with $H'=H$ ($H'=iH$) for cases with ``$PT$" or ``psH" (``$CP$" or ``CS"). 
     In this table, $c$ denotes the codimension of EP$n$s. Namely, in $D$ dimensions ($D \geq c$), EP$n$s form a ($D-c$)-dimensional manifold.
     For systems of $m$ bands ($m>n$), the characterization is done by analyzing the effective $n\times n$-Hamiltonian which describes $n$-band touching. 
     }
    \label{tab: W}
\end{table}

\section{
EP3s in four dimensions}
\label{sec: EP3 4D}

We illustrate our arguments for generic EP3s which are robust against perturbations in four dimensions.
First, we introduce the resultant vector and the associated topological invariant, the resultant winding number. Then, we provide toy models for EP3s with arbitrary winding numbers. Finally, we provide a fermion doubling theorem for EP3s in a four-dimensional BZ.

\subsection{Resultants and their winding}

Since the essential features of an EP3 are captured by a $3\times 3$-Jordan block~\footnote{
In principle, an $n\times n$ Hamiltonian can be diagonalizable. This case, however, requires additional tuning of $\mathcal O(n^2)$ parameters; the generic form of a degenerate Hamiltonian is a Jordan block.
}, we consider a generic non-Hermitian Hamiltonian of a $3\times 3$-matrix $H(\bm{k})$ in the four-dimensional space parameterized by $\bm{k}=(k_1,\cdots,k_4)^T$. The corresponding eigenvalues are obtained as roots of the characteristic polynomial $P(\lambda,\bm{k})=\mathrm{det}[H(\bm{k})-\lambda \1]$ ($\lambda \in C$) with $\1$ being the $3 \times 3$ identity matrix. 
Here, $P(\lambda)$ is a polynomial of degree three.
We suppose that three bands touch at $\bm{k}_0$ in four dimensions with energy $\epsilon_0\in\mathbb{C}$. In this case, the characteristic polynomial has a triple root $P(\lambda,\bm{k}_0)=(\lambda-\epsilon_0)^3$. The triple root is captured by vanishing resultants $r_j=0$ ($j=1,2$), where $r_j$ is defined as
~\footnote{
The resultant of two polynomials is defined as the determinant of the Sylvester matrix, a matrix whose elements are coefficients of these polynomials.
The resultant vanishes when these polynomials have a common root\!
}
\begin{equation}
    \label{eq: Res 3band}
    r_j(\bm{k}) = 
    \mathrm{Res}\left[
    \partial^{2-j}_\lambda
     P(\lambda,\bm{k}),
    \partial^2_\lambda
     P(\lambda,\bm{k})
    \right]
\end{equation}
(for more details, see Sec.~\ref{sec: R=0 vs EPn} of SM~\cite{Suppl}).
In other words, a generic EP3 is captured by the resultant vector
\begin{equation}
    \label{eq: Rvec NoSym}
\bm{R}(\bm{k})=
\Big(\re{[r_1]},\im{[r_1]},
\re{[r_2]},\im{[r_2]}\Big)^{T},
\end{equation}
which vanishes at EP3s in four dimensions.
This argument indicates that the codimension of generic EP3s is $c=4$, meaning that 
generic EP3s forms a $D-4$ manifold in a $D$-dimensional parameter space ($D\geq 4$).

To quantify the stability of generic EP3s, we introduce the resultant winding number. 
We consider a three-dimensional sphere surrounding an EP3 in the momentum space and specify a point on the sphere by $\bm{p}=(p_1,p_2,p_3)^T$. On this sphere, the resultant vector remains finite and can thus be normalized to $\bm{n}=\bm{R}/\sqrt{\bm{R}\cdot \bm{R}}$. 
This normalized vector induces a map $\bm{n}$: $S^3\to S^3$ with $\bm{p}\mapsto \bm{n}(\bm{p})$, which may possess nontrivial topology classified as 
\begin{equation}
 \pi_{3}(S^3) =\mathbb{Z}
\end{equation}
(see Sec.~4.1 of Ref.~\cite{AHatcher_Textbook2022}).
The topology is characterized by winding number $W_3$~\footnote{
The topological invariant $W_3$ can be regarded as an extension of the one-dimensional winding number of the discriminant that characterizes the Abelian topology of EP2s~\cite{Yang_EPdoubling_PRL2021}.
}
~\footnote{
This resultant approach cannot be directly applied to Hermitian band-touching. This is because the resultant does not include information of wavefunctions whose topology protects Hermitian band-touching.
For instance, 2-band touching corresponds to vanishing discriminant $\mathrm{Res}[P(\lambda), \partial_\lambda P(\lambda)]=0$. However, a Weyl node, Hermitian 2-band touching, cannot be captured by the winding of the discriminant
because it is always real and non-negative\!
}
\begin{eqnarray} \label{eq:3DWN}
W_3 &=& \frac{\epsilon^{ijkl}}{2\pi^2} \int \! d^3\bm{p} \, f_{ijkl}, \\
f_{ijkl} &=& n_i \partial_1 n_j \partial_2 n_k \partial_3 n_l, \nonumber 
\end{eqnarray}
with the anti-symmetric tensor $\epsilon^{ijkl}$ satisfying $\epsilon^{1234}=1$.
Here, $\partial_\mu$ ($\mu=1,2,3$) denotes the derivative with respect to $p_\mu$.

Reference~\cite{Tang_EPn_Science2020} reports that 
a coupled acoustic cavity system hosts an EP3 in four dimensions whose topological characterization remains unclear.
Computing the resultant winding number elucidates that this EP3 is characterized by $W_3=1$ which explicitly clarifies its stability. The details of the system are provided in Sec.~\ref{sec: EP3 phonon} of SM~\cite{Suppl}. 
\subsection{
Toy models of generic EP3
}
\label{sec: EP3 4D toy}

\begin{figure}[t]
\centering
\def\svgwidth{\linewidth}
\begingroup%
  \makeatletter%
  \providecommand\color[2][]{%
    \errmessage{(Inkscape) Color is used for the text in Inkscape, but the package 'color.sty' is not loaded}%
    \renewcommand\color[2][]{}%
  }%
  \providecommand\transparent[1]{%
    \errmessage{(Inkscape) Transparency is used (non-zero) for the text in Inkscape, but the package 'transparent.sty' is not loaded}%
    \renewcommand\transparent[1]{}%
  }%
  \providecommand\rotatebox[2]{#2}%
  \newcommand*\fsize{\dimexpr\f@size pt\relax}%
  \newcommand*\lineheight[1]{\fontsize{\fsize}{#1\fsize}\selectfont}%
  \ifx\svgwidth\undefined%
    \setlength{\unitlength}{255.11811024bp}%
    \ifx\svgscale\undefined%
      \relax%
    \else%
      \setlength{\unitlength}{\unitlength * \real{\svgscale}}%
    \fi%
  \else%
    \setlength{\unitlength}{\svgwidth}%
  \fi%
  \global\let\svgwidth\undefined%
  \global\let\svgscale\undefined%
  \makeatother%
  \begin{picture}(1,0.51111111)%
    \lineheight{1}%
    \setlength\tabcolsep{0pt}%
    \put(-0.00833378,0.47877323){\color[rgb]{0,0,0}\makebox(0,0)[lt]{\lineheight{1.25}\smash{\begin{tabular}[t]{l}(a)   $z_2=0$\end{tabular}}}}%
    \put(0.43769378,0.15169963){\color[rgb]{0,0,0}\rotatebox{38}{\makebox(0,0)[rt]{\lineheight{1.25}\smash{\begin{tabular}[t]{r}Re$z_1$\end{tabular}}}}}%
    \put(0.02553688,0.32439584){\color[rgb]{0,0,0}\rotatebox{-85}{\makebox(0,0)[lt]{\lineheight{1.25}\smash{\begin{tabular}[t]{l}Re$E$\end{tabular}}}}}%
    \put(0.01037252,0.38581882){\color[rgb]{0,0,0}\makebox(0,0)[rt]{\lineheight{1.25}\smash{\begin{tabular}[t]{r}$1$\end{tabular}}}}%
    \put(0.02424361,0.14363437){\color[rgb]{0,0,0}\makebox(0,0)[rt]{\lineheight{1.25}\smash{\begin{tabular}[t]{r}-$1$\end{tabular}}}}%
    \put(0.01730807,0.25884696){\color[rgb]{0,0,0}\makebox(0,0)[rt]{\lineheight{1.25}\smash{\begin{tabular}[t]{r}$0$\end{tabular}}}}%
    \put(0.04695176,0.10294499){\color[rgb]{0,0,0}\makebox(0,0)[rt]{\lineheight{1.25}\smash{\begin{tabular}[t]{r}-$1$\end{tabular}}}}%
    \put(0.12941172,0.04685449){\color[rgb]{0,0,0}\makebox(0,0)[rt]{\lineheight{1.25}\smash{\begin{tabular}[t]{r}$0$\end{tabular}}}}%
    \put(0.23401814,-0.01093996){\color[rgb]{0,0,0}\makebox(0,0)[rt]{\lineheight{1.25}\smash{\begin{tabular}[t]{r}$1$\end{tabular}}}}%
    \put(0.25370844,0.01554158){\color[rgb]{0,0,0}\rotatebox{-30}{\makebox(0,0)[rt]{\lineheight{1.25}\smash{\begin{tabular}[t]{r}Im$z_1$\end{tabular}}}}}%
    \put(0.45160298,0.12011592){\color[rgb]{0,0,0}\makebox(0,0)[lt]{\lineheight{1.25}\smash{\begin{tabular}[t]{l}-$1$\end{tabular}}}}%
    \put(0.36193869,0.05458798){\color[rgb]{0,0,0}\makebox(0,0)[lt]{\lineheight{1.25}\smash{\begin{tabular}[t]{l}$0$\end{tabular}}}}%
    \put(0.27227438,-0.01093996){\color[rgb]{0,0,0}\makebox(0,0)[lt]{\lineheight{1.25}\smash{\begin{tabular}[t]{l}$1$\end{tabular}}}}%
    \put(0,0){\includegraphics[width=\unitlength,page=1]{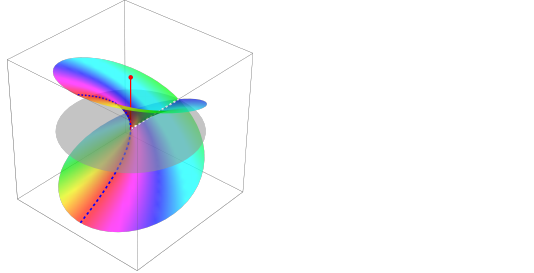}}%
    \put(0.92907483,0.15169982){\color[rgb]{0,0,0}\rotatebox{38}{\makebox(0,0)[rt]{\lineheight{1.25}\smash{\begin{tabular}[t]{r}Re$z_2$\end{tabular}}}}}%
    \put(0.48304723,0.47877323){\color[rgb]{0,0,0}\makebox(0,0)[lt]{\lineheight{1.25}\smash{\begin{tabular}[t]{l}(b)   $z_1=0$\end{tabular}}}}%
    \put(0.51723558,0.32356114){\color[rgb]{0,0,0}\rotatebox{-85}{\makebox(0,0)[lt]{\lineheight{1.25}\smash{\begin{tabular}[t]{l}Re$E$\end{tabular}}}}}%
    \put(0.74508949,0.01554149){\color[rgb]{0,0,0}\rotatebox{-30}{\makebox(0,0)[rt]{\lineheight{1.25}\smash{\begin{tabular}[t]{r}Im$z_2$\end{tabular}}}}}%
    \put(0.50237151,0.38581882){\color[rgb]{0,0,0}\makebox(0,0)[rt]{\lineheight{1.25}\smash{\begin{tabular}[t]{r}$1$\end{tabular}}}}%
    \put(0.5162426,0.14363437){\color[rgb]{0,0,0}\makebox(0,0)[rt]{\lineheight{1.25}\smash{\begin{tabular}[t]{r}-$1$\end{tabular}}}}%
    \put(0.50930706,0.25884696){\color[rgb]{0,0,0}\makebox(0,0)[rt]{\lineheight{1.25}\smash{\begin{tabular}[t]{r}$0$\end{tabular}}}}%
    \put(0.51485257,0.10294499){\color[rgb]{0,0,0}\makebox(0,0)[lt]{\lineheight{1.25}\smash{\begin{tabular}[t]{l}$1$\end{tabular}}}}%
    \put(0.61945898,0.04685449){\color[rgb]{0,0,0}\makebox(0,0)[rt]{\lineheight{1.25}\smash{\begin{tabular}[t]{r}$0$\end{tabular}}}}%
    \put(0.7240654,-0.01093996){\color[rgb]{0,0,0}\makebox(0,0)[rt]{\lineheight{1.25}\smash{\begin{tabular}[t]{r}-$1$\end{tabular}}}}%
    \put(0.94298401,0.12011592){\color[rgb]{0,0,0}\makebox(0,0)[lt]{\lineheight{1.25}\smash{\begin{tabular}[t]{l}$1$\end{tabular}}}}%
    \put(0.85331963,0.05458798){\color[rgb]{0,0,0}\makebox(0,0)[lt]{\lineheight{1.25}\smash{\begin{tabular}[t]{l}$0$\end{tabular}}}}%
    \put(0.76365526,-0.01093996){\color[rgb]{0,0,0}\makebox(0,0)[lt]{\lineheight{1.25}\smash{\begin{tabular}[t]{l}-$1$\end{tabular}}}}%
    \put(0,0){\includegraphics[width=\unitlength,page=2]{fig1.pdf}}%
    \put(0.97775448,0.03401947){\color[rgb]{0,0,0}\rotatebox{90}{\makebox(0,0)[lt]{\lineheight{1.25}\smash{\begin{tabular}[t]{l}Im$E$\end{tabular}}}}}%
    \put(0.97942676,0.48264351){\color[rgb]{0,0,0}\makebox(0,0)[rt]{\lineheight{1.25}\smash{\begin{tabular}[t]{r}$1$\end{tabular}}}}%
    \put(0.97942676,0.00164931){\color[rgb]{0,0,0}\makebox(0,0)[rt]{\lineheight{1.25}\smash{\begin{tabular}[t]{r}-$1$\end{tabular}}}}%
    \put(0.97942676,0.24214641){\color[rgb]{0,0,0}\makebox(0,0)[rt]{\lineheight{1.25}\smash{\begin{tabular}[t]{r}$0$\end{tabular}}}}%
    \put(0,0){\includegraphics[width=\unitlength,page=3]{fig1.pdf}}%
  \end{picture}%
\endgroup%
\caption{
Cuts of the spectrum of the Hamiltonian~(\ref{eq: toy W3=-1 EP3 no symm}) showcasing the EP$3$ at the origin. The real (imaginary) part of the eigenvalues is represented
as height (color). 
Fermi arcs, i.e., coincidences of two eigenvalues' real (imaginary) parts are shown as white (blue) dashed lines.
The panels~(a)~and~(b) show the parameter subspace of \(z_2=0\) and  \(z_1=0\), respectively.
}	
\label{fig: Ek EP3in4D}
\end{figure}

To explicitly demonstrate the emergence of a topologically protected EP3 in four dimensions, we consider the non-Hermitian Hamiltonian 
\begin{eqnarray}
\label{eq: toy W3=-1 EP3 no symm}
H(\bm{k}) &=&
\left(
\begin{array}{ccc}
 0    &  1     & 0 \\
 0    &  0     & 1 \\
z_2   & z_1    & 0
\end{array}
\right),
\end{eqnarray}
with $z_1=k_1+ik_2$ and $z_2=k_3+ik_4$.
As shown in Fig.~\ref{fig: Ek EP3in4D}, this model hosts an EP3 at $\bm{k}_0=0$ in four dimensions.
The resultant vector of this model is computed as 
\begin{eqnarray}
\label{eq: Rvec EP3 in 4D toy}
\bm{R}(\bm{k}) &=& 36 \Big( k_1, k_2, 6k_3, 6k_4 \Big),
\end{eqnarray}
for computation of a general case, see Sec.~\ref{sec: nxn toy model} of SM~\cite{Suppl}. 
Evaluating the integral of Eq.~\eqref{eq:3DWN}, we obtain the winding number $W_3=1$. This result is to be expected since the map defined in Eq.~\eqref{eq: Rvec EP3 in 4D toy} is obtained from the identity map via rescaling and the restriction to the sphere.

Models with a higher value of $W_3$ are obtained by taking powers of $z$'s. For instance, the replacement $z_1\to z^2_1=k^2_1-k^2_2 +2ik_1k_2$ yields a toy model with $W_3=2$. Similarly, complex conjugation $z_1=k_1+i k_2 \to z^*_1=k_1-i k_2$ in Eq.~\eqref{eq: toy W3=-1 EP3 no symm} yields a toy model with $W_3=-1$.

\begin{figure}[b]
\def\svgwidth{\linewidth}
\begingroup%
  \makeatletter%
  \providecommand\color[2][]{%
    \errmessage{(Inkscape) Color is used for the text in Inkscape, but the package 'color.sty' is not loaded}%
    \renewcommand\color[2][]{}%
  }%
  \providecommand\transparent[1]{%
    \errmessage{(Inkscape) Transparency is used (non-zero) for the text in Inkscape, but the package 'transparent.sty' is not loaded}%
    \renewcommand\transparent[1]{}%
  }%
  \providecommand\rotatebox[2]{#2}%
  \newcommand*\fsize{\dimexpr\f@size pt\relax}%
  \newcommand*\lineheight[1]{\fontsize{\fsize}{#1\fsize}\selectfont}%
  \ifx\svgwidth\undefined%
    \setlength{\unitlength}{246.61417323bp}%
    \ifx\svgscale\undefined%
      \relax%
    \else%
      \setlength{\unitlength}{\unitlength * \real{\svgscale}}%
    \fi%
  \else%
    \setlength{\unitlength}{\svgwidth}%
  \fi%
  \global\let\svgwidth\undefined%
  \global\let\svgscale\undefined%
  \makeatother%
  \begin{picture}(1,0.27586207)%
    \lineheight{1}%
    \setlength\tabcolsep{0pt}%
    \put(0,0){\includegraphics[width=\unitlength,page=1]{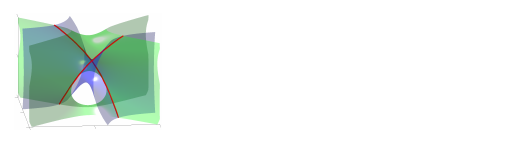}}%
    \put(0.3076002,0.00565551){\color[rgb]{0,0,0}\makebox(0,0)[lt]{\lineheight{1.25}\smash{\begin{tabular}[t]{l}$2$\end{tabular}}}}%
    \put(0.05332969,0.00322986){\color[rgb]{0,0,0}\makebox(0,0)[lt]{\lineheight{1.25}\smash{\begin{tabular}[t]{l}-$2$\end{tabular}}}}%
    \put(0.18366594,0.00272196){\color[rgb]{0,0,0}\makebox(0,0)[lt]{\lineheight{1.25}\smash{\begin{tabular}[t]{l}$0$\end{tabular}}}}%
    \put(0.03155643,0.09567568){\color[rgb]{0,0,0}\makebox(0,0)[rt]{\lineheight{1.25}\smash{\begin{tabular}[t]{r}-$2$\end{tabular}}}}%
    \put(0.03155643,0.23609945){\color[rgb]{0,0,0}\makebox(0,0)[rt]{\lineheight{1.25}\smash{\begin{tabular}[t]{r}$2$\end{tabular}}}}%
    \put(0.03065241,0.16375041){\color[rgb]{0,0,0}\makebox(0,0)[rt]{\lineheight{1.25}\smash{\begin{tabular}[t]{r}$0$\end{tabular}}}}%
    \put(0.04298266,0.02267283){\color[rgb]{0,0,0}\makebox(0,0)[rt]{\lineheight{1.25}\smash{\begin{tabular}[t]{r}$2$\end{tabular}}}}%
    \put(0.03169542,0.04611173){\color[rgb]{0,0,0}\makebox(0,0)[rt]{\lineheight{1.25}\smash{\begin{tabular}[t]{r}$0$\end{tabular}}}}%
    \put(0.02040819,0.06955026){\color[rgb]{0,0,0}\makebox(0,0)[rt]{\lineheight{1.25}\smash{\begin{tabular}[t]{r}-$2$\end{tabular}}}}%
    \put(0.24452104,0.00367189){\color[rgb]{0,0,0}\makebox(0,0)[lt]{\lineheight{1.25}\smash{\begin{tabular}[t]{l}$k_2$\end{tabular}}}}%
    \put(0.02839088,0.01764194){\color[rgb]{0,0,0}\makebox(0,0)[rt]{\lineheight{1.25}\smash{\begin{tabular}[t]{r}$k_1$\end{tabular}}}}%
    \put(0.03065241,0.20308884){\color[rgb]{0,0,0}\makebox(0,0)[rt]{\lineheight{1.25}\smash{\begin{tabular}[t]{r}$k_3$\end{tabular}}}}%
    \put(0.03548676,0.24502205){\color[rgb]{0,0,0}\makebox(0,0)[lt]{\lineheight{1.25}\smash{\begin{tabular}[t]{l}(a)\end{tabular}}}}%
    \put(0,0){\includegraphics[width=\unitlength,page=2]{fig2.pdf}}%
    \put(0.62419058,0.05102234){\color[rgb]{0,0,0}\makebox(0,0)[lt]{\lineheight{1.25}\smash{\begin{tabular}[t]{l}-$2$\end{tabular}}}}%
    \put(0.5426855,0.01363629){\color[rgb]{0,0,0}\makebox(0,0)[lt]{\lineheight{1.25}\smash{\begin{tabular}[t]{l}$2$\end{tabular}}}}%
    \put(0.59523497,0.03163542){\color[rgb]{0,0,0}\makebox(0,0)[lt]{\lineheight{1.25}\smash{\begin{tabular}[t]{l}$0$\end{tabular}}}}%
    \put(0.53209129,0.00506892){\color[rgb]{0,0,0}\makebox(0,0)[rt]{\lineheight{1.25}\smash{\begin{tabular}[t]{r}$2$\end{tabular}}}}%
    \put(0.45087784,0.02114298){\color[rgb]{0,0,0}\makebox(0,0)[rt]{\lineheight{1.25}\smash{\begin{tabular}[t]{r}$0$\end{tabular}}}}%
    \put(0.37102048,0.03766904){\color[rgb]{0,0,0}\makebox(0,0)[rt]{\lineheight{1.25}\smash{\begin{tabular}[t]{r}-$2$\end{tabular}}}}%
    \put(0.56587368,0.01950382){\color[rgb]{0,0,0}\makebox(0,0)[lt]{\lineheight{1.25}\smash{\begin{tabular}[t]{l}$k_1$\end{tabular}}}}%
    \put(0.49707176,0.01150383){\color[rgb]{0,0,0}\makebox(0,0)[rt]{\lineheight{1.25}\smash{\begin{tabular}[t]{r}$k_2$\end{tabular}}}}%
    \put(0.37197332,0.24502205){\color[rgb]{0,0,0}\makebox(0,0)[lt]{\lineheight{1.25}\smash{\begin{tabular}[t]{l}(b)\end{tabular}}}}%
    \put(0.3656489,0.06798928){\color[rgb]{0,0,0}\makebox(0,0)[rt]{\lineheight{1.25}\smash{\begin{tabular}[t]{r}-$2$\end{tabular}}}}%
    \put(0.36610093,0.21157716){\color[rgb]{0,0,0}\makebox(0,0)[rt]{\lineheight{1.25}\smash{\begin{tabular}[t]{r}$2$\end{tabular}}}}%
    \put(0.36610093,0.13955721){\color[rgb]{0,0,0}\makebox(0,0)[rt]{\lineheight{1.25}\smash{\begin{tabular}[t]{r}$0$\end{tabular}}}}%
    \put(0.36610093,0.17692306){\color[rgb]{0,0,0}\makebox(0,0)[rt]{\lineheight{1.25}\smash{\begin{tabular}[t]{r}$k_3$\end{tabular}}}}%
    \put(0,0){\includegraphics[width=\unitlength,page=3]{fig2.pdf}}%
    \put(0.96483701,0.01784542){\color[rgb]{0,0,0}\makebox(0,0)[lt]{\lineheight{1.25}\smash{\begin{tabular}[t]{l}-$2$\end{tabular}}}}%
    \put(0.76902903,0.00359868){\color[rgb]{0,0,0}\makebox(0,0)[lt]{\lineheight{1.25}\smash{\begin{tabular}[t]{l}$2$\end{tabular}}}}%
    \put(0.87380939,0.01168091){\color[rgb]{0,0,0}\makebox(0,0)[lt]{\lineheight{1.25}\smash{\begin{tabular}[t]{l}$0$\end{tabular}}}}%
    \put(0.70224947,0.09542653){\color[rgb]{0,0,0}\makebox(0,0)[rt]{\lineheight{1.25}\smash{\begin{tabular}[t]{r}-$2$\end{tabular}}}}%
    \put(0.70224947,0.23210555){\color[rgb]{0,0,0}\makebox(0,0)[rt]{\lineheight{1.25}\smash{\begin{tabular}[t]{r}$2$\end{tabular}}}}%
    \put(0.70224947,0.1600856){\color[rgb]{0,0,0}\makebox(0,0)[rt]{\lineheight{1.25}\smash{\begin{tabular}[t]{r}$0$\end{tabular}}}}%
    \put(0.75959068,0.01514036){\color[rgb]{0,0,0}\makebox(0,0)[rt]{\lineheight{1.25}\smash{\begin{tabular}[t]{r}-$2$\end{tabular}}}}%
    \put(0.7265009,0.04721826){\color[rgb]{0,0,0}\makebox(0,0)[rt]{\lineheight{1.25}\smash{\begin{tabular}[t]{r}$0$\end{tabular}}}}%
    \put(0.7024395,0.06719613){\color[rgb]{0,0,0}\makebox(0,0)[rt]{\lineheight{1.25}\smash{\begin{tabular}[t]{r}$2$\end{tabular}}}}%
    \put(0.91947973,0.01443639){\color[rgb]{0,0,0}\makebox(0,0)[lt]{\lineheight{1.25}\smash{\begin{tabular}[t]{l}$k_1$\end{tabular}}}}%
    \put(0.71934524,0.02973158){\color[rgb]{0,0,0}\makebox(0,0)[rt]{\lineheight{1.25}\smash{\begin{tabular}[t]{r}$k_2$\end{tabular}}}}%
    \put(0.70224947,0.19790345){\color[rgb]{0,0,0}\makebox(0,0)[rt]{\lineheight{1.25}\smash{\begin{tabular}[t]{r}$k_3$\end{tabular}}}}%
    \put(0.70961802,0.24502205){\color[rgb]{0,0,0}\makebox(0,0)[lt]{\lineheight{1.25}\smash{\begin{tabular}[t]{l}(c)\end{tabular}}}}%
  \end{picture}%
\endgroup%
\caption{
A generic EP3 visualized as a crossing point of the concomitant lines of EP2s in three dimensions. 
The green (blue) surfaces denote the real (imaginary) part of the discriminant of the characteristic polynomial of the model in Eq.~\eqref{eq: toy W3=-1 EP3 no symm}. The data for $k_4=-1, 0, 1$ are displayed in panels (a), (b), and (c), respectively. 
The lines of EP2s appear as band touching between different pairs of bands. 
In panel (b), the two lines of EP2s intersect with each other at an EP3, making a transition from over-crossing to under-crossing or vice versa.
}
\label{fig: k4-slice EP3in4D}
\end{figure}

Finally, we give a prescription for locating EP3s in a three-dimensional context because an EP3 does not necessarily emerge at the specific point $\bm{k}=0$. EP3s are generally surrounded by more abundant EP2s. Fixing one parameter, e.g., $k_4$, one is left with a three-dimensional model $H(\bm{k})$ that generally contains lines of EP2s. The EP3 can then be regarded as a crossing point of two such lines, on each of which different pairs of bands touch  [see Fig.~\ref{fig: k4-slice EP3in4D}]. Such a crossing point can be clearly identified in a parametric sweep of $k_4$.

\subsection{
Doubling theorem
}
\label{sec: EPn nosym doubling}

Having introduced the three-dimensional resultant winding number, we derive a concomitant doubling theorem for EP3s in three-band models, which is 
a direct consequence of the 
Poincar{\'e}-Hopf theorem~\cite{Mathai_doubling_CommMath2017,Mathai_doubling_JPhysA2017}.
We suppose that there exists an arbitrary number of discrete EP3s labeled by $I=1,2,\ldots$ in a four-dimensional Brillouin zone (BZ). Then, the summation of all their corresponding winding numbers vanishes $\sum_I W_{3I}=0$, since this sum is computed as
\begin{eqnarray}
\label{eq: W3 no symm doubling}
\sum_I \epsilon^{ijkl} \! \int_{S^3_I} \! d^3\bm{p} \, f_{ijkl}  &=& \epsilon^{ijkl}\! \int_{\partial \mathrm{BZ}} \! d^3\bm{p} \, f_{ijkl}=0,
\end{eqnarray}
up to a common prefactor. 
Here, $S^3_I$ denotes a three-dimensional sphere surrounding $I$-th EP3, and the ``boundary"~\footnote{
Mathematically, what is commonly denoted as the BZ boundary \(\partial\)BZ really is a double cover of a choice of \(n-1\)-skeleton of the \(n\)-dimensional BZ\!
} 
of the four-dimensional BZ is denoted by $\partial$BZ. The second equal sign is obtained from the periodicity and the orientability of the BZ.
This constitutes the doubling theorem: an EP3 with $W_3=1$ must be accompanied by an EP3 with $W_3=-1$ in the BZ.

\section{
Generic EP$n$ in $2n-2$ dimensions
}
\label{sec: EPn nosym W}

The above argument is straightforwardly generalized to EP$n$s in $2n-2$ dimensions.
We describe general $n$-fold band touching by an $n\times n$ non-Hermitian Hamiltonian $H(\bm{k})$. From the characteristic polynomial $P(\lambda,\bm{k})=\mathrm{det}[H(\bm{k})-\lambda \1]$ of degree $n$, we obtain the resultant vector of $2n-2$ components as 
\begin{eqnarray}
\label{eq: Rvec NoSym 2n-2}
\bm{R}(\bm{k})
\!\! &=& \!\!
\Big(\re{[r_1]},\im{[r_1]}, \cdots,
\re{[r_{n-1}]},\im{[r_{n-1}]}\Big)^{T}, 
\\
r_j(\bm{k})
\!\!&=&\!\! 
\label{eq: Res}
\mathrm{Res}[
\partial^{(n-1-j)}_\lambda P(\lambda,\bm{k}),
\partial^{(n-1)}_\lambda P(\lambda,\bm{k})
].
\end{eqnarray}
An $n$-tiple root of $P(\lambda)$ corresponds to the vanishing resultant vector, $\bm{R}(\bm{k}_0)=0$, which amounts to $2n-2$ constraints. 
Thus, the codimension of generic EP$n$s is $c=2n-2$. Namely,
a generic EP$n$ in $2n-2$ dimensions is a point stable against perturbations.
In $D$ dimensions ($D>2n-2$), EP$n$s form a $(D-2n+2)$-dimensional manifold~\cite{Delplace_Resul_PRL21,Mandal_EP3_PRL21}.

As is the case of EP3s in four dimensions, EP$n$s in $2n-2$ dimensions ($n=2,3,4,\ldots$) may possess nontrivial topology of the map from the $(2n-3)$-dimensional sphere around the EP$n$ in the parameter space to the ($2n-2$)-dimensional normalized resultant vector 
$\bm{n}=\bm{R}/\sqrt{\bm{R}\cdot \bm{R}}$. 
Such maps are topologically classified as (see Sec.~4.1 of Ref.~\cite{AHatcher_Textbook2022})
\begin{eqnarray}
\pi_{2n-3}(S^{2n-3}) &=& \mathbb{Z}
\end{eqnarray}
where the invariant is characterized by winding number $W_{2n-3}$
\begin{eqnarray}
\label{eq: W(2n-3) nosymm}
W_{2n-3} &=& \frac{ \epsilon^{i_1\cdots i_{2n-2}} }{ A_{2n-3} } 
\int \! d^{2n-3}\bm{p} \, f_{i_1\cdots i_{2n-2}}, \\
f_{i_1\cdots i_{2n-2}} &=& n_{i_1} \partial_1 n_{i_2} \partial_2 n_{i_3} \cdots \partial_{2n-3} n_{i_{2n-2}}, \nonumber \\
A_{2n-3} &=& \frac{(n-2)!}{2 \pi^{(n-1)}}, \nonumber
\end{eqnarray}
with the anti-symmetric tensor $\epsilon^{i_1\cdots i_{2n-2}}$ satisfying $\epsilon^{12\cdots 2n-2}=1$.
The vector $\bm{p}=(p_1,p_2,\ldots,p_{2n-3})^T$ specifies a point on a ($2n-3$)-dimensional sphere which encloses the EP$n$ in the momentum space.
The resultant winding number $W_{2n-3}$
topologically characterizes a manifold of generic EP$n$s whose codimension is $2n-2$.

Generalizing Eq.~\eqref{eq: toy W3=-1 EP3 no symm}, we provide an $n$-band toy model hosting a generic EP$n$. The Hamiltonian reads
\begin{eqnarray}
\label{eq: Jordan ptb main}
H(\bm{k})&=&
\left(
\begin{array}{cccc}
0 & 1& & 0 \\
\vdots & \ddots& \ddots&  \\
0 & & 0&  1\\
z_{n-1} & \cdots& z_1& 0
\end{array}
\right)
\end{eqnarray}
with $z_{j}=k_{2j-1}+i k_{2j}$ ($j=1,\ldots,n-1$). 
The resultant vector of this toy model is obtained as (see Sec.~\ref{sec: nxn toy model} of SM~\cite{Suppl}) 
\begin{eqnarray}
R_{2j-1} &=& \alpha_{n,j} k_{2j-1}, \\
R_{2j} &=& \alpha_{n,j} k_{2j},
\end{eqnarray}
with $\alpha_{n,j} = (-1)^{j(n+1)}(n!)^{j+1}(n-1-j)! ~$. 
Thus, this lone EP$n$ at $\bm{k}=0$ is characterized by the winding number $W_{2n-3}=1$. 
Replacing $z_1\to (k_1\pm i k_2)^m$ yields a toy model hosting an EP$n$ with $W_{2n-3}=\pm m$ ($m=1,2,\ldots$), which we explicitly calculate in Sec.~\ref{sec: EPn nosymm W=m} of SM~\cite{Suppl}. 

The ($2n-3$)-dimensional resultant winding number $W_{2n-3}$ leads to the doubling theorem of generic EP$n$s for an arbitrary $n$-band model in the $(2n-2)$-dimensional BZ. In a similar way to the case of $n=3$ (see Sec.~\ref{sec: EPn nosym doubling}), the sum of $W_{2n-3}$ is deformed into an integral over the ``boundary" of the BZ ($\partial$BZ). The integral over $\partial$BZ vanishes due to the periodicity and the orientability of the BZ. 
This fact  $\sum_I W_{2n-3I}=0$ constitutes the doubling theorem: an EP$n$ with $W_{2n-3}=1$ must be 
accompanied by an EP$n$ with $W_{2n-3}=-1$ in the BZ.

\section{
Symmetry-protected EP$n$ in $n-1$ dimensions
}
\label{sec: EPn PTsymm}

The above topological characterization of EP$n$s can be extended to systems with symmetry. We emphasize that the following argument for systems with $PT$-symmetry applies directly to systems with pseudo-Hermiticity, $CP$-symmetry, or chiral symmetry.

A $PT$-symmetric Hamiltonian satisfies 
\begin{eqnarray}
\label{eq: PT symm}
U_{\mathrm{PT}} H^*(\bm{k}) U^\dagger_{\mathrm{PT}}
&=&
H(\bm{k}),
\end{eqnarray}
where ``*" denotes complex conjugation, and the unitary matrix $U_{\mathrm{PT}}$ satisfies $U_{\mathrm{PT}}U^*_{\mathrm{PT}}=\1$. The above constraint implies that the corresponding characteristic polynomial $P(\lambda,\bm{k})$ takes the form 
\begin{eqnarray}
\label{eq: P(lambda)=P^*(lambda)}
P(\lambda)&=&\mathrm{det}[U_{PT} H^*(\bm{k})U^\dagger_{PT}-\lambda\1] \nonumber\\ 
&=&\mathrm{det}[H^*(\bm{k})-\lambda\1]
\nonumber\\ 
&=&[P(\lambda^*)]^*.
\end{eqnarray}
Namely, $P(\lambda)$ is a polynomial with real coefficients. 
Its $n$-tiple root is captured by the vanishing resultant vector defined as
\begin{eqnarray}
\label{eq: Rvec PT}
 \bm{R} &=& (r_1,r_2,\ldots,r_{n-1})^T, \\
\label{eq: Res PT}
r_j(\bm{k})&=&\mathrm{Res}[\partial^{n-1-j}_{\lambda}P,\partial^{n-1}_{\lambda} P],
\end{eqnarray}
with $j=1,2,\ldots,n-1$.
We note that for polynomials with real coefficients, the resultants $r_j(\bm{k})$ are real.
Therefore, $\bm{R}=0$ amounts to $n-1$ real constraints, which means that 
the codimension of the symmetry-protected EP$n$s is $c=n-1$. Namely, 
symmetry-protected EP$n$s with $PT$-symmetry are stable against
perturbations preserving the relevant symmetry 
in $n-1$ dimensions. 
In $D$ dimensions ($D>n-1$), EP$n$s form a $(D-n+1)$-dimensional manifold~\cite{Delplace_Resul_PRL21,Mandal_EP3_PRL21}.

The symmetry-protected EP$n$s in $n-1$ dimensions $(n=2,3,4,\ldots)$ may possess nontrivial topology of the map from the $(n-2)$-dimensional sphere around the EP$n$ in the parameter space to the ($n-1$)-dimensional normalized resultant vector $\bm{n}=\bm{R}/\sqrt{\bm{R}\cdot \bm{R}}$. Such maps are topologically classified as (see Sec.~4.1 of Ref.~\cite{AHatcher_Textbook2022})
\begin{eqnarray}
\pi_{n-2}(S^{n-2}) &=& \mathbb{Z},
\end{eqnarray}
where the invariant is characterized by the winding number $W_{n-2}$ 
\begin{eqnarray}
\label{eq: W PT}
W_{n-2} &=& \frac{ \epsilon^{i_1\cdots i_{n-1} } }{ A_{n-2} } 
\int \! d^{n-2}\bm{p} \, f_{i_1\cdots i_{n-1}}, \\
f_{i_1\cdots  i_{n-1}} &=& n_{i_1} \partial_1 n_{i_2} \partial_2 n_{i_3} \cdots \partial_{n-2} n_{i_{n-1}}, \nonumber 
\end{eqnarray}
with the area of the $(n-2)$-dimensional sphere $A_{n-2}$ given by 
\begin{eqnarray}
A_{2m-1} &=& \frac{2\pi^m}{(m-1)!}, \\
A_{2m-2} &=& \frac{2^{2m-1} \pi^{m-1} (m-1)! }{(2m-2)!}, \nonumber 
\end{eqnarray}
for $m=1,2,\ldots~$.
The vector $\bm{p}=(p_1,p_2,\ldots,p_{n-2})^T$ specifies a point on a ($n-2$)-dimensional sphere which encloses the EP$n$ in the momentum space.
The resultant winding number $W_{n-2}$
topologically characterizes a manifold of symmetry-protected EP$n$s whose codimension is $n-1$.

We provide an $n$-band toy model hosting a symmetry-protected EP$n$. The Hamiltonian reads
\begin{eqnarray}
\label{eq: Jordan ptb PT}
H(\bm{k})&=&
\left(
\begin{array}{cccc}
0 & 1& & 0 \\
\vdots & \ddots& \ddots&  \\
0 & & 0&  1\\
z_{n-1} & \cdots& z_1& 0
\end{array}
\right),
\end{eqnarray}
with $z_j=k_j$ [$z_j=(-1)^j k_j$] ($j=1,2,\ldots,n$) for odd [even] $n$.
This Hamiltonian is an $n\times n$-matrix with real entries, satisfying $PT$-symmetry
with $U_{PT}=\1$ [see Eq.~\eqref{eq: PT symm}].
As shown in Sec.~\ref{sec: nxn toy model} of SM~\cite{Suppl}, 
the resultant vector is given by 
\begin{eqnarray}
 R_j &=& |\alpha_{n,j}| k_j, \\
\alpha_{n,j} &=& (-1)^{j(n+1)}(n!)^{j+1}(n-1-j)!~, \nonumber 
\end{eqnarray}
with real numbers $\alpha_{n,j}$.
Thus, the symmetry-protected EP$n$ emerging in the toy model [Eq.~\eqref{eq: Jordan ptb PT}] is characterized by $W_{n-2}=1$. The replacement $z_1+i z_2 \to (q_1 \pm i q_2)^m$ yields a toy model that hosts an EP$n$ with $W_{n-2}=\pm m$ ($m=1,2,\ldots$). Here, $q$'s are defined as $q_j=k_j$ [$q_j=(-1)^j k_j$] for odd [even] $n$.

The ($n-2$)-dimensional resultant winding number leads to the doubling theorem of EP$n$s for an arbitrary $PT$-symmetric $n$-band model in the $(n-1)$-dimensional BZ.
In a similar way to the case of generic EP3s [see Eq.~\eqref{eq: W3 no symm doubling}], the sum of $W_{n-2}$ is written as an integral over the ``boundary" of the BZ ($\partial$BZ). This integral over $\partial$BZ vanishes due to the periodicity and the orientability of the BZ. This fact $\sum_I W_{n-2I}=0$ constitutes the doubling theorem: an EP$n$ with $W_{n-2}=1$ must be accompanied by an EP$n$ with $W_{n-2}=-1$ in the BZ.

We finish this section by extensions to systems with pseudo-Hermiticity, $CP$-symmetry, or chiral symmetry.
Pseudo-Hermiticity is written as 
\begin{eqnarray}
\label{eq: psH Hami}
U_{\mathrm{pH}} H(\bm{k}) U^\dagger_{\mathrm{pH}}&=& H^\dagger(\bm{k}),
\end{eqnarray}
for a unitary matrix satisfying $U^2_{\mathrm{pH}}=\1$. Under pseudo-Hermiticity, the characteristic polynomial $P(\lambda)$ ($\lambda\in\mathbb{R}$) satisfies Eq.~\eqref{eq: P(lambda)=P^*(lambda)}. Hence the argument for the $PT$-symmetric case applies directly.

$CP$-symmetry and chiral symmetry are respectively written as
\begin{eqnarray}
\label{eq: CP Hami}
U_{CP} H^*(\bm{k}) U^\dagger_{CP}&=& -H(\bm{k}), \\
\label{eq: chiral Hami}
U_{\mathrm{c}} H^\dagger(\bm{k}) U^\dagger_{\mathrm{c}}&=& -H(\bm{k}),
\end{eqnarray}
where $U_{CP}$ ($U_{\mathrm{c}}$) is a unitary matrix satisfying $U_{CP}U^*_{CP}=\1$ ($U^2_{\mathrm{c}}=\1$). By the identification $H'(\bm{k})=i H(\bm{k})$, Eq.~\eqref{eq: CP Hami} [Eq.~\eqref{eq: chiral Hami}] 
coincides with the constraint of $PT$-symmetry Eq.~\eqref{eq: P(lambda)=P^*(lambda)} [pseudo-Hermiticity Eq.~\eqref{eq: psH Hami}].
Hence the argument of the $PT$-symmetric case applies to both $CP$- and chiral symmetry as well.

\section{
Discussion
}
\label{sec: summary}
In this paper, we have systematically characterized the Abelian topology of generic EP$n$s and symmetry-protected EP$n$s in arbitrary dimensions [see Table~\ref{tab: W}]. 
Specifically, introducing resultant winding numbers, we have characterized the topology of generic EP$n$s in $2n-2$ dimensions as well as symmetry-protected EP$n$s in $n-1$ dimensions with $PT$-, $CP$-, pseudo-Hermiticity or chiral symmetry. 
In a $D$-dimensional parameter space with $D\geq c$, the resultant winding number characterize
a $(D-c)$-dimensional manifold of generic [symmetry-protected] EP$n$s, whose codimension is $c=2n-2$ [$c=n-1$].
The introduced winding number leads to concomitant doubling theorems for both generic and symmetry-protected EP$n$s appearing in $n$-band models.

We note that 
Ref.~\cite{Tang_EP3Mech_NatComm2023} also modified the resultant winding number for $PT$-symmetry-protected EP3 introduced in Ref.~\cite{Delplace_Resul_PRL21}.
The novelty of our results is further generalizing the resultant winding numbers for both generic and symmetry-protected EP$n$ for arbitrary $n$.
These resultant winding numbers are verified by constructing toy models of generic and symmetry-protected EP$n$ for arbitrary $n$. 
These results demonstrate the usefulness of the choice of resultants defined in Eqs.~\eqref{eq: Rvec NoSym 2n-2} and \eqref{eq: Rvec PT} which also differentiates from the previous work~\cite{Tang_EP3Mech_NatComm2023}. 

Our systematic characterization of the topological nature of EP$n$s increases the fundamental understanding of non-Hermitian multiband structures of direct physical importance. 
The stability of generic EP$3$s in four dimensions will be of importance in non-Hermitian Floquet systems, where periodic driving effectively induces a fourth dimension. 
Going beyond Bloch Hamiltonians, recent studies show that multifold EPs scaling with system sizes have been reported in, e.g., the Hatano-Nelson model, paving the way to realize topologically stable EP$n$s for any $n$ by increasing system sizes. 
Concerning symmetry-protected EP$n$s, our topological classification has important consequences already in two and three dimensions. 
While the previous work has characterized symmetry-protected EPs up to third order, our general argument applies to pertinent EP4s in three dimensions.
Our result makes their topological classification directly relevant to three-dimensional photonic crystals, where $PT$ symmetry can be implemented as a perfect balance between gain and loss.
Furthermore, both symmetry-protected EP$3$s and EP$4$s have been experimentally realized in nitrogen-vacancy spin systems~\cite{Wu_EP3openQ_NatNano2024}, single-photon setups~\cite{Wang_EPnPTPhotoSciAdv2023}  and correlated quantum many-body systems~\cite{Crippa_EP4Corr_PRB2021}, showcasing the breadth of applications in numerous physical systems. 

Beyond experimental implications, one natural continuation from a theoretical point of view is to investigate the interplay between the Abelian and non-Abelian topology in the characterization of EP$n$s. 
We expect non-Abelian contributions to appear when additional bands are added, i.e., when EP$n$s are studied in $m$-band models, for some integer $m>n$. Such a classification scheme is expected to include braiding of the complex eigenvalues around EP$n$s. 

The Abelian framework outlined here guarantees the topological stability of EP$n$s and
serves as a robust cornerstone in the continued unraveling and understanding of the intriguing properties of the non-Hermitian topological band theory.

\begin{acknowledgments}
T.Y. thanks Pierre Delplace for collaboration in the previous work~\cite{Delplace_Resul_PRL21}.
T.Y. also thanks Guancong Ma for fruitful discussions of an acoustic system.
M.S. acknowledges fruitful discussions with Anton Montag.
This work is supported by JSPS KAKENHI Grant Nos.~JP21K13850, JP23KK0247, JSPS Bilateral Program No.~JPJSBP120249925, and the Swedish Research Council (grant 2018-00313), the Wallenberg Academy Fellows program of the Knut and Alice Wallenberg Foundation (grant 2018.0460) and the G\"oran Gustafsson Foundation for Research in Natural Sciences and Medicine. L.R. is supported by the Knut and Alice Wallenberg Foundation under Grant No.~2017.0157.
M.S. is supported by the Swedish Research Council (VR) under grant No.~2024-00272.
T.Y. is grateful for the support from the ETH Pauli Center for Theoretical Studies and the Grant from Yamada Science Foundation. 
\end{acknowledgments}


%

\clearpage

\renewcommand{\thesection}{S\arabic{section}}
\setcounter{section}{0}
\renewcommand{\theequation}{S\arabic{equation}}
\setcounter{equation}{0}
\renewcommand{\thefigure}{S\arabic{figure}}
\setcounter{figure}{0}
\renewcommand{\thetable}{S\arabic{table}}
\setcounter{table}{0}
\makeatletter
\c@secnumdepth = 2
\makeatother

\onecolumngrid
\begin{center}
 {\large \textmd{Supplemental Materials:} \\[0.3em]
 {\bfseries 
 Winding Topology of Multifold Exceptional Points
 }
 }
\end{center}

\setcounter{page}{1}


\section{EP$n$ and the vanishing resultants}
\label{sec: R=0 vs EPn}
For an $n\times n$-Hamiltonian, the vanishing resultants
indicate the emergence of an EP$n$.
Specifically, we prove that the characteristic polynomial is written as 
\begin{eqnarray}
\label{eq: P sim (lambda-E0)^n}
P(\lambda)&=& (-1)^n(\lambda-\epsilon_0)^n
\end{eqnarray}
with $\epsilon_0 \in \mathbb{C}$
if and only if
\begin{eqnarray}
\label{eq: all Res =0}
\mathrm{Res}[P(\lambda), \partial^{n-1}_\lambda P(\lambda) ] &=& 0,\nonumber \\
\mathrm{Res}[\partial^2_\lambda P(\lambda), \partial^{n-1}_\lambda P(\lambda) ] &=& 0, \nonumber \\
&\vdots& \nonumber \\
\mathrm{Res}[\partial^{n-2}_\lambda P(\lambda), \partial^{n-1}_\lambda P(\lambda) ] &=& 0 
\end{eqnarray}
are satisfied.

\subsection{
A proof of
Eq.~(\ref{eq: P sim (lambda-E0)^n})
$\Rightarrow$
Eq.~(\ref{eq: all Res =0})
}
When Eq.~(\ref{eq: P sim (lambda-E0)^n})
holds, we have
\begin{eqnarray}
\partial^l_\lambda P(\lambda)
&=& (-1)^{n}
\frac{n!}{(n-l)!} (\lambda- \epsilon_0)^{n-l}
\end{eqnarray}
with $l=0,1,\ldots,n-1$,
indicating that $\partial^l_\lambda P(\lambda)$ and 
$\partial^{l'}_\lambda P(\lambda)$
have the common root 
$\lambda=\epsilon_0$ for $l,l'=0,1,\ldots,n-1$.
Thus, Eq.~(\ref{eq: all Res =0}) holds.

\subsection{
A proof of
Eq.~(\ref{eq: all Res =0})
$\Rightarrow$
Eq.~(\ref{eq: P sim (lambda-E0)^n})
}
We consider the characteristic polynomial 
$P(\lambda)$ of degree $n$.
Its $(n-1)$-th derivative is linear, \(\partial_\lambda^{n-1}P = c( \lambda - \epsilon_0 )\) with some constant \(c\), and thus has only one root \(\lambda=\epsilon_0\). 
The vanishing resultants in Eq.~\eqref{eq: all Res =0} imply that \(P(\lambda)\) and all of its derivatives share a root with this linear polynomial. 
Since there is only one such root, \(P\) and all of its derivatives must have \(\epsilon_0\) as the root.  
For a polynomial of finite degree, this is only possible if there are no other roots, hence \(P\) has an $n$-tiple root at \(\epsilon_0\).

Recalling that the leading coefficient of the characteristic polynomial is \((-1)^{n}\), we can fix the overall prefactor and obtain $P(\lambda)=(-1)^n(\lambda -\epsilon_0)^n$.

\section{
EP3 in a coupled acoustic cavity system
}
\label{sec: EP3 phonon}

Reference~\onlinecite{Tang_EPn_Science2020} reports that a coupled acoustic cavity system hosts EP3 in a four-dimensional parameter space. This section provides the details of its characterization based on our resultant winding number.

The coupled acoustic cavity system is described by the following effective Hamiltonian
\begin{eqnarray}
H&=&
(\omega_0+i\gamma_0)\1
+\kappa
\left(
\begin{array}{ccc}
i\sqrt{2}(1+\Lambda) & -1 & 0 \\
-1 & i\Xi & -1\\
0 & -1 & -i\sqrt{2}(1+\Lambda)
\end{array}
\right),
\end{eqnarray}
with 
$\Xi=\delta_{\mathrm{f}}+i\delta_{\mathrm{A}}$
and 
$\Lambda=\delta_{\mathrm{g}}+i\delta_{\mathrm{B}}$
($\delta_{\mathrm{f}}, \, \delta_{\mathrm{f}}, \,  \delta_{\mathrm{A}}, \,\delta_{\mathrm{B}}\in \mathbb{R}$).
Here, $\delta_{\mathrm{A}}$ and $\delta_{\mathrm{B}}$ ($\delta_{\mathrm{g}}$ and $\delta_{\mathrm{f}}$) are detuning (gain or loss) of each cavity. Coupling between cavities is described by $\kappa$ ($\kappa \in \mathbb{R}$).
The onsite eigenfrequency and intrinsic loss are described by $\omega_0$ and $\gamma_0$ ($\omega_0,\gamma_0\in \mathbb{R}$), respectively.
The experiment is carried out for $\omega_0=19,613 \:\mathrm{rad/s}$
and $\kappa=49.9 \: \mathrm{rad/s}$.
The EP3 is characterized by analyzing the matrix in the second term scaled by $\kappa$.

The EP3 emerges at the origin of a four-dimensional space parameterized by $(\delta_{\mathrm{f}},\delta_{\mathrm{A}},\delta_{\mathrm{g}},\delta_{\mathrm{B}})$.
Numerically computing the resultant winding number [Eq.~(4)
] yields $W_3=1$.

\section{Resultant of a toy model of EP$n$}
\label{sec: nxn toy model}

We consider the following Hamiltonian of an $n\times n$-matrix
\begin{eqnarray}
\label{eq: Jordan ptb}
H&=&
\left(
\begin{array}{cccc}
0 & 1& & 0 \\
\vdots & \ddots& \ddots&  \\
0 & & 0&  1\\
z_{n-1} & \cdots& z_1& 0
\end{array}
\right)
\end{eqnarray}
with $z_j$ ($j=1,\ldots,n-1$) being functions of $\bm{k}$.
The characteristic polynomial 
$P_{n}(\lambda)=\mathrm{det}[H-\1 \lambda]$
is given by
\begin{eqnarray}
\label{eq: Pn(lambda)}
P_n(\lambda)&=&
(-1)^{n}\Big[ \lambda^n- (z_1\lambda^{n-2}+z_2\lambda^{n-3}+  \cdots+z_{n-1-l}\lambda^{l} +\cdots +z_{n-1})
\Big].\nonumber
\end{eqnarray}
The above relation can be seen by noting the recurrence relation 
\begin{eqnarray}
\label{eq: Pn(lambda) recurr}
P_n(\lambda)&=& -\lambda P_{n-1}(\lambda) +(-1)^{n-1} z_{n-1},
\end{eqnarray}
and 
\begin{eqnarray}
\label{eq: P2(lambda)}
P_2(\lambda)&=& \lambda^2-z_1.
\end{eqnarray}
We note that
$(n-1)$-th derivative of $P_n(\lambda)$ is given by $\partial^{(n-1)}_\lambda P_n(\lambda)=(-1)^n n! \lambda$.

Let us consider two polynomials
\begin{eqnarray}
\label{eq: f(x)}
f(x)&=& a_m x^m+a_{m-1} x^{m-1}+\cdots+a_0, \\
g(x)&=& b_1 x,
\end{eqnarray}
with coefficients $a_i, i=1,\ldots,m$ and $b_1$.
The resultant of these polynomials is given by 
\begin{eqnarray}
\label{eq: Res [f g]}
\mathrm{Res}[f(x),g(x)] &=& (-1)^{m} b^{m}_1 a_0,
\end{eqnarray}
which can be seen from a direct computation
\begin{eqnarray}
\mathrm{Res}[f(x),g(x)] 
&=& \mathrm{det}
\left[
\begin{array}{ccccc}
a_m & a_{m-1}& \cdots &a_1& a_0 \\
b_1 &  0     & \cdots &0 & 0  \\
0   &  b_1   & \cdots &0 & 0  \\
0   &  0     & \ddots &\vdots & \vdots \\
\vdots   &        & 0 &b_1 & 0 \\
\end{array}
\right].
\end{eqnarray}

We apply this fact to the characteristic polynomial in Eq.~\eqref{eq: Pn(lambda)} with
$\partial^{n-1}_\lambda P_n(\lambda)=(-1)^n n! \lambda$.
Using Eq.~(\ref{eq: Res [f g]}), we obtain 
\begin{eqnarray}
\mathrm{Res}[\partial^{l}_\lambda P_n(\lambda)  ,\partial^{n-1}_\lambda P_n(\lambda)] &=&
(-1)^{(n+1)(l+1)}(n!)^{n-l} l! z_{n-1-l},
\nonumber \\
\end{eqnarray}
which leads to
\begin{eqnarray}
r_j &=& \alpha_{n,j} z_{j}, \nonumber  \\
\alpha_{n,j} &=& (-1)^{j(n+1)}(n!)^{j+1}(n-1-j)!~.
\end{eqnarray}

Therefore, specifying the momentum dependence of $z_j$, we obtain the specific form of the resultant vector.

For $z_j=k_{2j-1}+ik_{2j}$, the toy model of a general EP$n$, we obtain the resultant vector from the real and imaginary parts of \(r_j\) as
\begin{eqnarray}
R_{2j-1} &=& \alpha_{n,j} k_{2j-1}, \\
R_{2j} &=& \alpha_{n,j} k_{2j},
\end{eqnarray}
following Eq.~(\ref{eq: Rvec NoSym 2n-2}).

For $z_j=(-1)^{j(n+1)}k_j$, the toy model of symmetry-protected EP$n$s, we construct the resultant vector from the real parts of \(r_j\) as
\begin{eqnarray}
R_{j} &=& |\alpha_{n,j}| k_j,
\end{eqnarray}
following Eq.~\eqref{eq: Rvec PT}.

\section{Toy model taking an arbitrary $W_{2n-3}$}
\label{sec: EPn nosymm W=m}

We prove that the EP$n$ in toy model~(\ref{eq: Jordan ptb})
is characterized by 
$W_{2n-3}=\pm m$ $(m=0,1,2,\dots)$ for
\begin{eqnarray}
z_1 &=& 
\begin{cases}
    k_1^2+k_2^2 & \text{if \(m=0\)}
    \\
    \textcolor{black}{(k_1 \pm i k_2)^{m}} & \text{otherwise},
    \end{cases}
    \\
z_j &=& k_{2j-1}+i k_{2j} \quad (j=2,\ldots,n-1).
\end{eqnarray}

\subsection{
Case of positive $W_{2n-3}$
}

Firstly, we note that the winding number is computed from
\begin{eqnarray}
\label{eq: W_{2n-3} dom}
W_{2n-3}=\sum_{\bm{n}(\bm{k}_l)=\bm{n}_0}
\mathrm{sgn} [J(\bm{k}_l)]
\end{eqnarray}
after fixing some arbitrary regular point $\bm{n}_0$ in the target space $S^{2n-3}$.
The sum is over the preimage of this point, i.e.,
over $\bm{k}_l$ satisfying $\bm{n}(\bm{k}_l)=\bm{n}_0$, and $\sgn(x)$ takes $1$ ($-1$) for $x>0$ ($x<0$).
The Jacobian $J(\bm{k})$ is given by
\begin{eqnarray}
J(\bm{k}) &=&
\left|
\frac{\partial(n_1,n_2,\ldots,n_{2n-2})}
{\partial(k_1,k_2,\ldots,k_{2n-2})}
\right|.
\end{eqnarray}

By making use of Eq.~(\ref{eq: W_{2n-3} dom}), we compute the winding number.
We take as the point \(\bm n_0\) either the north or the south pole of the sphere, according to $r$'s giving 
$\bm{n}_0=\mathrm{sgn}(\alpha_{n1})(1,0,\cdots,0)^T$.
In this case, resultants
\begin{eqnarray}
r_{j} &=& 
\left\{
\begin{array}{ll}
\alpha_{n1} , &\quad j=1  \\
0 , &\quad j=2,\ldots,n-1~,
\end{array}
\right.
\end{eqnarray}
corresponds to 
constraints on \(z_j\) 
\begin{eqnarray}
z_{j} &=&  
\left\{
\begin{array}{ll}
1 , &\quad j=1  \\
0 , &\quad j=2,\ldots,n-1~.
\end{array}
\right.
\end{eqnarray}
Thus, for 
\begin{eqnarray}
\label{eq: z1 = k1+ik2 m}
z_1=(k_1+ik_2)^m, \quad\quad z_j=k_{2j-1}+ik_{2j} 
\end{eqnarray}
with $m=1,2,\ldots$ and $j=2,\ldots,n-1$, we find the preimages in terms of the 
the $m$-th root of unity 
\begin{eqnarray}
k_{l1}+ik_{l2}=\mathrm{e}^{i\frac{2\pi l}{m}}, \quad
k_{l3}=\cdots=k_{l2n-2}=0, 
\end{eqnarray}
with $l=0,\ldots,m-1$.

From Eq.~(\ref{eq: z1 = k1+ik2 m}), we have
\begin{eqnarray}
\frac{\partial z_1}{\partial k_1} &=& m \mathrm{e}^{-i\frac{2\pi l}{m}}, \nonumber \\
\frac{\partial z_1}{\partial k_2} &=& im \mathrm{e}^{-i\frac{2\pi l}{m}},
\end{eqnarray}
for $\bm{k}=\bm{k}_l$.

Thus, for each $l$, we obtain the Jacobian $J(\bm{k}_l)$ from the real and imaginary parts of \(z_i\) as
\begin{eqnarray}
&&
J(\bm{k}_l)
 = \mathrm{det}
\left[
\begin{array}{cccc}
J_{2\times 2} & 0 &  \cdots & 0  \\
0 & 1 & &  \\
\vdots & & \ddots&  \\
0 & & & 1
\end{array}
\right], 
\\
&&
J_{2\times 2} = 
\left(
\begin{array}{cc}
m\cos(\frac{2\pi l}{m}) & -m \sin(\frac{2\pi l}{m}) \\
m \sin(\frac{2\pi l}{m}) &  m\cos(\frac{2\pi l}{m})
\end{array}
\right).
\end{eqnarray}
Thus, the sign of this Jacobian is always $+1$ for arbitrary $l$, which leads to
\begin{eqnarray}
W_{2n-3}=\sum_{\bm{n}(\bm{k})=\bm{n}_0}
\mathrm{sgn} [J(\bm{k})] =m
\end{eqnarray}
for $m=1,2,\ldots$~.

\subsection{
Case of negative $W_{2n-3}$
}

In a similar way, we can compute the winding number $W_{2n-3}$ of negative values.
The difference is the Jacobian. Because we have
\begin{eqnarray}
\frac{\partial z_1}{\partial k_1} &=& m \mathrm{e}^{-i\frac{2\pi l}{m}}, \nonumber \\
\frac{\partial z_1}{\partial k_2} &=& -im \mathrm{e}^{-i\frac{2\pi l}{m}},
\end{eqnarray}
for $\bm{k}=\bm{k}_l$ ($l=0,1,\ldots,m-1$), we obtain
\begin{eqnarray}
&&
J(\bm{k}_l)
 = \mathrm{det}
\left[
\begin{array}{cccc}
J_{2\times 2} & 0 &  \cdots & 0  \\
0 & 1 & &  \\
\vdots & & \ddots&  \\
0 & & & 1
\end{array}
\right], \\
&&
J_{2\times 2} = 
\left(
\begin{array}{cc}
m\cos(\frac{2\pi l}{m}) & -m \sin(\frac{2\pi l}{m}) \\
-m \sin(\frac{2\pi l}{m}) &  -m\cos(\frac{2\pi l}{m})
\end{array}
\right).
\end{eqnarray}
Thus, the sign of this Jacobian is $-1$ for arbitrary $l$, which leads to
\begin{eqnarray}
 W_{2n-3}&=&-m
\end{eqnarray}
with $m=1,2,\ldots$~.

\subsection{Case of zero $W_{2n-3}$}
For the model with \(m=0\), we note that the vector \(\bm R\) does not cover the full sphere, since \(\operatorname{Im}r_1(\bm k) = 0\). 
The homotopy invariant of a non-surjective map is always zero. This can be seen by applying the summation formula Eq.~(\ref{eq: W_{2n-3} dom}) to the empty preimage of any point with non-zero \(R_2=\operatorname{Im}r_1\).

\end{document}